\documentclass[12pt]{article}
\usepackage[english,german,french,polish]{babel}
\usepackage[T1]{fontenc}
\usepackage{amsfonts}

\begin{document}

\selectlanguage{english}

\baselineskip 0.76cm
\topmargin -0.6in
\oddsidemargin -0.1in

\let\ni=\noindent

\renewcommand{\thefootnote}{\fnsymbol{footnote}}

\newcommand{\SM}{Standard Model }

\pagestyle {plain}

\setcounter{page}{1}

\pagestyle{empty}

~~~

\begin{flushright}
IFT-- 06/17
\end{flushright}

\vspace{0.5cm}

{\large\centerline{\bf Towards a realistic neutrino mass formula{\footnote{Work supported in part by the Polish Ministry of Education and Science, grant 1 PO3B 099 29 (2005-2007).}}}} 

\vspace{0.4cm}

{\centerline {\sc Wojciech Kr\'{o}likowski}}

\vspace{0.3cm}

{\centerline {\it Institute of Theoretical Physics, Warsaw University}}

{\centerline {\it Ho\.{z}a 69,~~PL--00--681 Warszawa, ~Poland}}

\vspace{0.5cm}

{\centerline{\bf Abstract}}

\vspace{0.2cm}

A new option of parameter constraint is described for the recently proposed neutrino mass
formula involving primarily three free parameters. The option implies the vanishing of a part of 
mass formula for the lowest mass neutrino, the part which may -- in an intuitive model -- be 
identified with its formal "intrinsic selfenergy". However, its actual mass induced then by 
another part of mass formula is considerable. In this option, our neutrino mass formula {\it 
predicts} all three neutrino masses as $m_1 \sim 2.5\times 10^{-3}$ eV, $ m_2 \sim 9.3\times 
10^{-3}$ eV and $m_3 \sim 5.0\times 10^{-2}$ eV, when two experimental estimates $\Delta 
m^2_{21} \sim 8.0\times 10^{-5}\;{\rm eV}^2$ and $\Delta m^2_{32} \sim 2.4\times 10^{-3}\;{\rm 
eV}^2$ are applied as an input.

\vspace{0.5cm}

\ni PACS numbers: 12.15.Ff , 14.60.Pq  .

\vspace{0.6cm}

\ni September 2006  

\vfill\eject

~~~
\pagestyle {plain}

\setcounter{page}{1}

\vspace{0.1cm}

\ni {\bf 1. Introduction}

\vspace{0.3cm}

In a recent paper [1], we have proposed a universal shape of empirical mass formula for leptons and quarks. This shape has been somehow supported by an intuitive model of formal intrinsic interactions which might work within leptons and quarks [2]. Such a mass formula for three generations $i = 1,2,3$ of four kinds $f = l, \nu, u, d$ of fundamental fermions reads:

\begin{equation}
m_{f_i}  =  \mu^{(f)}\, \rho_i \left(N^2_i + \frac{\varepsilon^{(f)} -1}{N^2_i} - \xi^{(f)} \right) \;.
\end{equation}

In the case of active neutrinos $\nu_i = \nu_1,\nu_2, \nu_3$, this formula has been proposed for neutrino Dirac masses $m^{(D)}_{\nu_i} = m^{(D)}_{\nu_1}, m^{(D)}_{\nu_2}, m^{(D)}_{\nu_3}$, while the neutrino effective masses $m_{\nu_i} = m_{\nu_1}, m_{\nu_2}, m_{\nu_3}$ are conjectured to be induced by a simple seesaw mechanism

\begin{equation} 
m_{\nu_i} = - \frac{m_{\nu_i}^{(D)\,2}}{M_{\nu_i}} = - \frac{m_{\nu_i}^{(D)}}{\zeta} > 0
\end{equation}

\ni with a very large parameter $\zeta \equiv M_{\nu_i}/m_{\nu_i}^{(D)} > 0$ involving neutrino Majorana masses $M_{\nu_i}$ assumed as proportional to $m_{\nu_i}^{(D)}$. This has led to the proposal of neutrino mass formula

\begin{equation}
m_{\nu_i} = \mu^{(\nu)}_{\rm eff}  \rho_i \left[1-  \frac{1}{\xi^{(\nu)}} \left(N^2_i + \frac{\varepsilon^{(\nu)} -1}{N^2_i}\right)\right] \;,
\end{equation}

\ni where

\begin{equation}
\mu^{(\nu)}_{\rm eff} \equiv \frac{\mu^{(\nu)} \xi^{(\nu)}}{\zeta} \,.
\end{equation}

In Eqs. (1) and (3), we use the notation

\begin{equation}
N_i = 1,3,5 \;\;\;,\;\;\; \rho_i = \frac{1}{29} \,,\, \frac{4}{29} \,,\, \frac{24}{29} 
\end{equation}

\ni ($\sum_i \rho_i = 1$), while $\mu^{(f)} > 0, \;\varepsilon^{(f)}$ and $\xi^{(f)} > 0$ denote three free parameters for any kind  $f = l, \nu, u, d$ of fundamental fermions ($f = \nu$ in the case of Eq. (3)). These free parameters are determined from three masses $m_{f_i} = m_{f_1}, m_{f_2}, m_{f_3}$ for any $f$ (if the masses are known) or {\it vice versa}. Thus, there are no numerical predictions for the masses, unless the free parameters are constrained.

In particular, for the precisely known case of charged leptons {$f = l$} it turns out that $\xi^{(l)} = 1.771\times 10^{-3} = 1.8\times 10^{-3} $, so $\xi^{(l)}$ is small. Putting approximately $\xi^{(l)} = 0$ as a constraint, we {\it predict}

\begin{equation} 
m_\tau = \frac{6}{125} (351m_\mu -  136 m_e) = 1776.80\;{\rm MeV} \,,
\end{equation}

\ni close to the experimental value $m_{\tau} = 1776.99^{+0.29}_{-0.26}\;{\rm MeV}$ [3]. Notice that in the case of $\xi^{(l)} = 0$  we determine

\begin{equation}
\mu^{(l)} = \frac{29 (9m_\mu - 4 m_e)}{320} = 85.9924\;{\rm MeV}\;,\; \varepsilon^{(l)} = \frac{320 m_e}{9 m_\mu - 4 m_e} = 0.172329 \,.
\end{equation}

\ni In Eqs. (6) and (7), the experimental values of $m_e$ and $m_\mu $ are used as an input. 

\vspace{0.3cm}

\ni {\bf 2. New constraint for neutrinos}

\vspace{0.3cm}

In the present paper, we discuss for neutrinos the option, where $\varepsilon^{(\nu)} = 0$ is put as a constraint, at least approximately. In such a case, the neutrino mass formula (3) can be rewritten as follows:

\begin{eqnarray}
m_{\nu_1} & = & \frac{\mu^{(\nu)}_{\rm eff}}{29} \,, \nonumber \\
m_{\nu_2} & = & \frac{\mu^{(\nu)}_{\rm eff}}{29}\; 4\left(1 -\frac{1}{\xi^{(\nu)}} \frac{80}{9}\right) \,, \nonumber \\
m_{\nu_3} & = & \frac{\mu^{(\nu)}_{\rm eff}}{29} \,24\left( 1 -\frac{1}{\xi^{(\nu)}} \frac{624}{25}\right) \,, 
\end{eqnarray}

\ni implying readily the neutrino mass sum rule

\begin{equation}
m_{\nu_3} = \frac{6}{125}\left(351 m_{\nu_2} - 904 m_{\nu_1}\right) \,.
\end{equation}

\ni This is different from the mass sum rule (6) valid for charged leptons in the case, where $\xi^{(l)} = 0$ is put approximately as a constraint.

Due to Eq. (9), we can write

\begin{equation}
\Delta m^2_{32}  +m^2_{\nu_2} = m^2_{\nu_3} = \left(\frac{6}{125}\right)^2 \left(351 m_{\nu_2} - 904 m_{\nu_1}\right)^2 \,,
\end{equation}

\ni where $\Delta m^2_{32}  \equiv m^2_{\nu_3} - m^2_{\nu_2} = \lambda(m^2_{\nu_2} - m^2_{\nu_1})$ with $\lambda \equiv \Delta m^2_{32}/\Delta m^2_{21}$ and $\Delta m^2_{21} \equiv m^2_{\nu_2} - m^2_{\nu_1}$. Hence, for $r \equiv m_{\nu_1}/m_{\nu_2}$, we obtain the following quadratic equation:  

\begin{equation}
\left[(904)^2 +\left(\frac{125}{6}\right)^2 \lambda\right] r^2 - 2 \left(351 \cdot 904\right) r + (351)^2 - \left(\frac{125}{6}\right)^2 (\lambda +1) =0 \,.
\end{equation}

\ni For the experimental estimates $\Delta m^2_{21} \sim 8.0\times 10^{-5}\;{\rm eV}^2$ and $\Delta m^2_{32} \sim 2.4\times 10^{-3}\;{\rm eV}^2$ [4] giving $\lambda \sim 30$, two solutions to Eq. (11) are 

\begin{equation}
r = \left\{ \begin{array}{l} 0.264 = 0.26 \\ 0.500 = 0.50 \end{array}\right. \,.
\end{equation}

 Then, choosing the smaller or larger of two solutions and making use of the experimental $\Delta m^2_{21}$ and $\Delta m^2_{32}$, we {\it predict} the neutrino masses

\begin{eqnarray}
m_{\nu_1} & \equiv & \sqrt {\frac{r^2 \Delta m^2_{21}}{1-r^2}} \sim \left\{ \begin{array}{l} 2.45\times 10^{-3}\;{\rm eV} = 2.5\times 10^{-3}\;{\rm eV} \\ 5.16\times 10^{-3}\;{\rm eV} = 5.2\times 10^{-3}\;{\rm eV} \end{array}\right.\,, \nonumber \\ m_{\nu_2} & \equiv & \sqrt {\frac{\Delta m^2_{21}}{1-r^2}} \sim \left\{ \begin{array}{l} 9.27\times 10^{-3}\;{\rm eV} =  9.3\times 10^{-3}\;{\rm eV} \\ 10.3\times 10^{-3}\;{\rm eV} = 10\times 10^{-3}\;{\rm eV} \end{array}\right.
\end{eqnarray}

\ni and

\vspace{-0.1cm}

\begin{equation}
m_{\nu_3} \equiv \sqrt{\Delta m^2_{32} + \frac{\Delta m^2_{21}}{1-r^2}} \sim \left\{\begin{array}{l} 4.99\times 10^{-2}\;{\rm eV} =  5.0\times 10^{-2}\;{\rm eV} \\ 5.01\times 10^{-2}\;{\rm eV} = 5.0\times 10^{-2}\;{\rm eV} \end{array}\right.
\end{equation}

\vspace{0.1cm}

\ni (the same value of $m_{\nu_3}$ follows, of course, from the mass sum rule (9)). Our actual numerical prediction is one of three values of $m_{\nu_i}$. From Eqs. (13) and (14) we get the prediction for the neutrino mass proportion 

\begin{equation}
m_{\nu_1} : m_{\nu_2} : m_{\nu_3} \sim \left\{\begin{array}{l} 1 : 3.8 : 20 = 0.26 : 1 : 5.4 = 0.049 : 0.19 : 1 \\ 1 :  2.0 : 9.7 = 0.50 : 1 : 4.8 = 0.10 : 0.21 : 1 \end{array}\right. \;.
\end{equation}

\vspace{0.1cm}

\ni Here, the predictions for neutrino masses are made possible by the imposed constraint $\varepsilon^{(\nu)} = 0$ and the input of experimental $\Delta m^2_{21}$ and $\Delta m^2_{32}$. We can see that in the option of $\varepsilon^{(\nu)} = 0$, the lowest neutrino mass $m_{\nu_1}$ is considerable {\it versus} $m_{\nu_2}$. Note that in this option, from Eqs. (8) we also determine 

\begin{equation}
\mu^{(\nu)}_{\rm eff} = 29/m_{\nu_1} \sim \left\{\begin{array}{l} 7.1\times 10^{-2}\;{\rm eV} \\ 15\times 10^{-2}\;{\rm eV} \end{array}\right. \;\;,\;\; \frac{1}{\xi^{(\nu)}} = \frac{9}{80}\left(1 - \frac{1}{4r} \right) \sim \left\{\begin{array}{l} 6.1\times 10^{-3} \\ 56\times 10^{-3} \end{array}\right. 
\end{equation}

\ni for the smaller or larger $r$.

\vspace{0.25cm}

\ni {\bf 3. Comparison of three options for neutrinos}

\vspace{0.25cm}

In Ref. [1], we considered as examples also the options of two other constraints, $\varepsilon^{(\nu)}/\xi^{(\nu)} = 1$ and $1/\xi^{(\nu)} = 0$. In the first case, the lowest neutrino mass $m_{\nu_1}$ vanishes. In the second, the ordering of 1 and 2 neutrino states is inverted, while the position of 3 neutrino state is normal.

A comparison of three different options of $\varepsilon^{(\nu)} = 0$, $\varepsilon^{(\nu)}/\xi^{(\nu)} = 1$ and $1/\xi^{(\nu)} = 0$ can be presented in the following listing:

\begin{center}
\begin{tabular}{c|c|c|c|c|c}
$\varepsilon^{(\nu)}/\xi^{(\nu)}$ & $1/\xi^{(\nu)}(10^{-3})$ & $\mu^{(\nu)}_{\rm eff} (10^{-2}\,{\rm eV})$ & $m_{\nu_1} (10^{-3}\,{\rm eV})$ & $m_{\nu_2} (10^{-3}\,{\rm eV})$ & $m_{\nu_3} (10^{-3}\,{\rm eV})$ \\ 
\hline & & & & & \\
0~ & 6.1 or 56 & 7.1 or 15 & 2.5 or 5.2 & 9.3 or 10 & 50 or 50 \\ 1~~ & 8.1 & 7.9 & 0~~ & 8.9 & 50 \\ -8.8~ & 0~~ & 4.5 & 15~~~ & 12~~~ & 51   
\end{tabular}
\end{center}

\vspace{0.2cm} 

\ni Here, the experimental estimates $|\Delta m^2_{21}| \sim 8.0\times 10^{-5}\,{\rm eV}^2$ and $\Delta m^2_{32} \sim 2.4\times 10^{-3}\,{\rm eV}^2$ are applied as an input.

\vspace{0.25cm}

\ni {\bf 4. An intrinsic interpretation}

\vspace{0.25cm}

It can be seen from the fundamental-fermion mass formula (1) that the parameter $\varepsilon^{(f)}$ appears as a factor in the formal "intrinsic selfenergy" ("intrinsic selfinteraction") of the fermion $f_1$ [2] {\it i.e.}, in a formal intrinsic quantity which may be identified with the first term in the mass formula of $f_1$: 

\begin{equation}
m_{f_1} = \frac{\mu^{(f)}}{29} \left(\varepsilon^{(f)} - \xi^{(f)} \right)\,.
\end{equation}

\ni Then, the second term may be interpreted as the formal "intrinsic binding energy"\, ("intrinsic binding interaction") of the fermion $f_1$ [2].

Thus, in the option of $\varepsilon^{(\nu)} = 0$, the formal "intrinsic selfenergy"\, vanishes for the lowest mass neutrino $\nu_1$, implying that the nonzero Dirac mass and effective mass of $\nu_1$ are  

\begin{equation}
m^{(D)}_{\nu_1} = -\frac{\mu^{(\nu)}}{29} \xi^{(\nu)}  \;\;\;\;{\rm and} \;\;\;\; m_{\nu_1} = \frac{\mu^{(\nu)}_{\rm eff}}{29} \,,
\end{equation}

\ni respectively, where the latter of them is induced by the former {\it via} a simple version of the seesaw mechanism (see Eqs. (2) and (3)).

In contrast, for the electron $e$, being the lowest charged lepton $l_1$, the formal "intrinsic selfenergy"\,is nonzero, as

\begin{equation}
m_{e} = \frac{\mu^{(l)}}{29} \left(\varepsilon^{(l)} - \xi^{(l)} \right) \simeq  \frac{\mu^{(l)}}{29} \varepsilon^{(l)}\,.
\end{equation}

\ni Here, $\varepsilon^{(l)} = 0.172329$ if $\xi^{(l)} = 1.771\times 10^{-3} = 1.8\times 10^{-3} $ is put approximately equal to zero (see Eq. (7)). 

The minimalization of formal "intrinsic selfenergy"\,(suggesting the vanishing of $\varepsilon^{(\nu)}$, at least approximately) is intuitively required for the lowest mass neutrino $\nu_1$, since -- in the \SM -- the active neutrinos are as much neutral (in gauge charges) as possible. This may be an intuitive argument for the option of $\varepsilon^{(\nu)} = 0$. So, our neutrino mass formula (3) involving primarily three free parameters seems to get a more realistic structure, when the option of parameter constraint $\varepsilon^{(\nu)} = 0$ is introduced (then, of course, it becomes also numerically predictive for neutrino masses as involving now only two free parameters).

\vspace{0.3cm}

\ni {\bf 5. Conclusions}

\vspace{0.3cm}

Concluding, we have described in this paper the option of $\varepsilon^{(\nu)} = 0$ for the neutrino mass formula (3) proposed previously in Ref. [1]. The constraint $\varepsilon^{(\nu)} = 0$ imposed on the free parameter $\varepsilon^{(\nu)}$, together with the input of experimental
$\Delta m^2_{21}$ and $\Delta m^2_{32}$, enables us to predict from the mass formula (3) all three neutrino masses $m_{\nu_1}, m_{\nu_2}, m_{\nu_3}$ and also to determine two remaining free parameters $\mu^{(\nu)}_{\rm eff}$ and $1/\xi^{(\nu)}$. In this option, the lowest neutrino mass $m_{\nu_1}$ is considerable: $m_{\nu_1}/m_{\nu_2} \sim 0.26$ or 0.50, while $ m_{\nu_2}/ m_{\nu_3} \sim 0.19$ or 0.21 and $m_{\nu_3} \sim 5.0\times 10^{-2}$ eV in both cases, if $\Delta m^2_{21} \sim 8.0\times 10^{-5}\,{\rm eV}^2$ and $\Delta m^2_{32} \sim 2.4\times 10^{-3}\,{\rm eV}^2$.

The vanishing of $\varepsilon^{(\nu)}$ may be connected with the vanishing of the formal "intrinsic selfenergy"\, for the lowest mass neutrino ${\nu_1}$. Then, the actual nonzero mass of $\nu_1$ may be interpreted as induced by the formal "intrinsic binding energy"\,  of $\nu_1$ {\it via} a simple version of the seesaw mechanism.

\vfill\eject

~~~~
\vspace{0.5cm}

{\centerline{\bf References}}

\vspace{0.5cm}

{\everypar={\hangindent=0.6truecm}
\parindent=0pt\frenchspacing

{\everypar={\hangindent=0.6truecm}
\parindent=0pt\frenchspacing

\vspace{0.2cm}

[1]~W. Kr\'{o}likowski, {\it Acta Phys. Pol.} {\bf B 37}, 2601 (2006) [{\tt hep--ph/0602018}].

\vspace{0.2cm}

[2]~W. Kr\'{o}likowski, {\tt hep--ph/0604148}; {\it cf.} also {\it Acta Phys. Pol.} {\bf B 33}, 2559 (2002) [{\tt hep--ph/0203107}].

\vspace{0.2cm}

[3]~W.M.~Yao {\it et al.} (Particle Data Group), {\it Review of Particle Physics, J.~Phys.} {\bf G 33}, 1 (2006).

\vspace{0.2cm}

[4]~{\it Cf. e.g.} G.L. Fogli, E. Lisi, A. Marrone, A. Palazzo, {\it Progr. Part. Nucl. Phys.} {\bf 57}, 742 (2006) [{\tt hep--ph/0506083}]; G.L. Fogli, E. Lisi, A. Mirizzi, D.~Montanino, P.D.~Serpico, {\tt hep--ph/0608321}.

\vspace{0.2cm}

\vfill\eject

\end{document}